\documentclass[notitlepage]{article}
\usepackage{graphicx}
\usepackage{amssymb,amsfonts,amsmath}
\usepackage{txfonts}
\usepackage[font=small]{caption}
\usepackage{xcolor}
\usepackage{setspace}
\usepackage{authblk}
\usepackage{fullpage}
\usepackage{url}
\usepackage{hyperref}

\def\lett#1{(\textbf{#1})}
\newcommand{\Prob}[1]{{\rm Pr}\left(#1\right)}

\begin{document}

\title{Shadow networks: Discovering hidden nodes with models of information flow}

\author[1,2,3,4]{James P.~Bagrow          \thanks{james.bagrow@uvm.edu}   }
\author[1,2,3,4]{Suma Desu                }
\author[1,2,3,4]{Morgan R.~Frank          }
\author[5,2,3]{Narine Manukyan            }
\author[1,2,3,4]{Lewis Mitchell           }
\author[1,2,3,4]{Andrew Reagan            }
\author[6]{Eric E.~Bloedorn               }
\author[6]{Lashon B.~Booker               }
\author[6]{Luther K.~Branting             }
\author[6]{Michael J.~Smith               }
\author[6,2,3,4]{Brian F.~Tivnan          }
\author[1,2,3,4]{Christopher M.~Danforth  }
\author[1,2,3,4]{Peter S.~Dodds           }
\author[5,2,3]{Joshua C.~Bongard          }
\affil[1]{\small Department of Mathematics \& Statistics, The University of Vermont, Burlington, VT 05401, USA}
\affil[2]{Vermont Complex Systems Center, The University of Vermont, Burlington, VT 05401, USA}
\affil[3]{Vermont Advanced Computing Core, The University of Vermont, Burlington, VT 05401, USA}
\affil[4]{Computational Story Lab, The University of Vermont, Burlington, VT 05401, USA}
\affil[5]{Department of Computer Science, The University of Vermont, Burlington, VT 05401, USA}
\affil[6]{The MITRE Corporation, McLean, VA 22102, USA}

\date{December 20, 2013}

\maketitle

\begin{abstract}
Complex, dynamic networks underlie many systems, and understanding these
networks is the concern of a great span of important scientific and engineering
problems. Quantitative description is crucial for this understanding yet, due
to a range of measurement problems, many real network datasets are incomplete.
Here we explore how accidentally missing or deliberately hidden nodes may be
detected in networks by the effect of their absence on predictions of the speed with
which information flows through the network.  
We use Symbolic Regression (SR) to learn models relating information flow to
network topology. These models show localized, systematic, and non-random
discrepancies when applied to test networks with intentionally masked nodes,
demonstrating the ability to detect the presence of missing nodes and where in
the network those nodes are likely to reside.
\end{abstract}

\section{Introduction}
\label{sec:intro}

The field of complex networks has emerged and matured over the last 15 years,
heralded by small-world~\cite{watts1998a} and scale-free
networks~\cite{barabasi1999a}, and principally enabled by the advent of readily
available large-scale datasets.  Much work has been focused on 
simple descriptions of complex networks, leading to an evolving collection of
structures, network
statistics~\cite{newman2003a,boccaletti2006a}, and generative
mechanisms~\cite{simon1955a,price1965a,barabasi1999a}.

All along, the problem of missing data has been both obvious and
ubiquitous---few network datasets are complete or nearly so---and
yet this issue has largely been ignored. The body of work that does exist on
missing data has mostly focused on the problem of unrecorded edges or
interactions~\cite{liben-nowell2007a,clauset_hierarchical_2008,wang_human_2011,bliss2013a}, 
while only some have explored the harder problems of node and context
omission~\cite{kossinets2006b,borgatti2006a,desilva2006a} using
various approaches such as inference based on maximum likelihood
estimation~\cite{maeno2009a,maeno2009b}.

While missing data is certainly understood to affect---sometimes
dramatically---different kinds of static network statistics in different
ways~\cite{kossinets2006b}, the effects of measurement error on dynamic, real
social networks~\cite{bagrow_collective_2011,dodds_temporal_2011,woolley-meza_eyjafjallajokull_2013,mitchell_geography_2013}
remain largely unknown. This problem is especially challenging when the amount
of data omission is not known and can only be estimated from the observed data
set. The implications for how to contend with a given network, suspected to
be corrupted in some fashion, are substantial. In the case of public health
policy, for example, positive evidence for the role of social contagion in
the spreading of such disparate attributes as happiness~\cite{fowler2008a},
obesity~\cite{christakis2007a}, and loneliness~\cite{cacioppo2009a}, have
been challenged due to their reliance on under-sampled reconstructed social
networks~\cite{shalizi_homophily_2011}.

A systematic framework to accommodate missing data for static and dynamic networks remains
elusive, and provides a great challenge to the network science community.  Much
success in the study of complex, dynamic networks has come from approaches born
out of statistical mechanics and dynamical systems, with the great example
arguably being Simon's rich-get-richer model underlying scale-free
networks~\cite{simon1955a,price1965a,barabasi1999a}.  Yet it is clear that many
adaptive complex systems are strongly algorithmic in nature, and are not well or
completely described by integrodifferential equations.

Briefly, our approach to studying missing or hidden node detection
is as follows. First, we construct a set of network topologies
(Sec.~\ref{subsec:networkmodels}). We then use an idealized transaction model to
simulate the flow of information ``packets'' across these networks. These packets
could represent IP packets flowing across a computer network, citations within a
scientific collaboration network, or messages passed among members of a social network such as
Twitter (Sec.~\ref{subsec:transactionmodel}). Next, the
resulting transaction data is collected and fed to a stochastic optimization
method. This goal of this step is to generate a mathematical model that predicts the
speed of information flow between pairs of nodes in the network, given
structural information about those nodes and the network they were drawn from
(Sec.~\ref{subsec:symbolicregression}). Finally, the evolved transaction model
is presented with rates of information flow between nodes from a different
network. If there are systematic errors or biases in the model's prediction of
information flow, this indicates that nodes may have been added or removed from
the network.

\begin{figure}[t!]
\centerline{\includegraphics[]{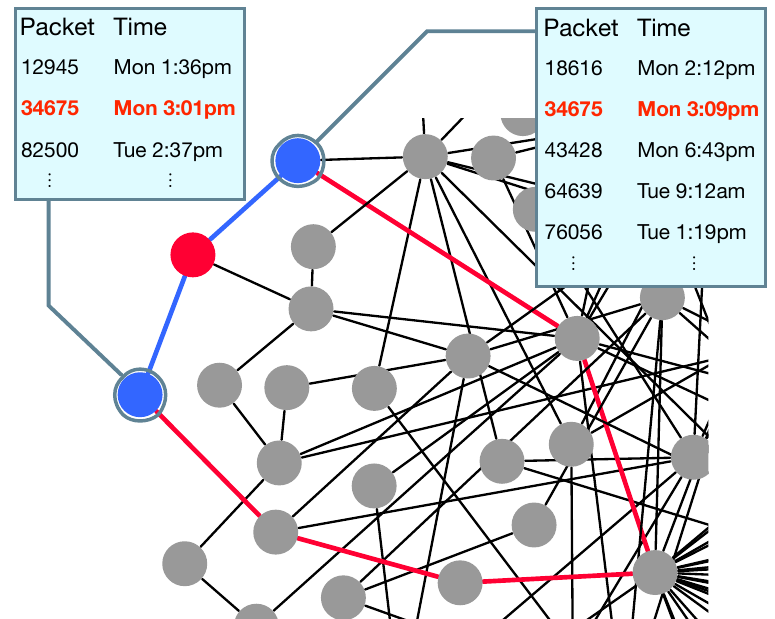}}
\caption{\label{fig:cartoonTransactionsModel}
    Illustration motivating the method. Nodes in this network pass information
    around (packets), and we monitor the arrival times of these packets. The two
    blue nodes appear much farther apart topologically when the red node is
    hidden. Given the observed information flows, the highlighted packet would appear to
    be arriving unusually quickly given the apparent long distance path it
    likely took (red links). This unexpectedly rapid flow may be a clue that unseen
    network elements are present.
}
\end{figure}

The intuition underlying our approach to node prediction may be clarified by
considering the cartoon example in Fig.~\ref{fig:cartoonTransactionsModel}. Two
nodes are connected by a third node, making them two steps apart on the network
topology. Due to their close proximity, information should flow between them
relatively quickly, on average. However, if the bridge node is hidden from us,
we may erroneously conclude these two nodes are actually quite far apart
(illustrated in the figure by the red path). We would then expect information
flow should be slow between them, even for information originating from other parties, and we would be surprised by the speed of flow
we actually observe. If we consistently overestimate the time it takes for
information to appear at one node after it appears at the other, then this provides
evidence that a hidden presence in the network is facilitating the flow of
information.

\section{Methods}
\label{sec:methods}

Here we describe the network topologies we will employ in this study, the
details of how we simulate information flow on these topologies, the predictive
models we generate for the flow times, and the test procedure and measurements
we use to explore how well hidden nodes can be detected.

\subsection{Network model}
\label{subsec:networkmodels}

To gauge the potential of our approach, we first developed it
against simulated transactional networks. We used scale-free
networks generated according to the common preferential attachment
model~\cite{simon1955a,price1965a,barabasi1999a}. Each scale-free network was
grown to a size of $N = 250$ nodes by adding new nodes one at a time, and
each new node attached to two existing nodes preferentially according to their
degree~\cite{barabasi1999a}. These undirected networks have a power-law degree
distribution $\Prob{k} \sim k^{-3}$. The earlier a node is added to the network,
the higher the degree it will tend to have. So \emph{hubs}, highly connected
nodes, tend to be among the early nodes of the network.

\subsection{Transaction dynamics}
\label{subsec:transactionmodel}

For each network that was constructed, we simulated transactions, the creation
and movement of \textbf{packets} of content, occurring between pairs of
connected nodes. Each packet carries a unique identifier so that it can be
tracked when it appears at different nodes in the network, and each node
maintains a growing, time-ordered list of the content packets it has received.
We simulated transactions  as follows. At each time step, each node is
activated with probability $p=1$. This may represent a member of an online
social network logging into their account, or a node in a computer network
being turned on. For each node that is activated, it \textbf{creates} a new
piece of content with probability $p_\mathrm{create}=1/9$ or \textbf{imports}
a piece of content from a neighbor with probability $p_\mathrm{import}=2
p_\mathrm{create}$. In the former case, this may correspond to a member of
the social network Twitter creating a new tweet; in the latter case it may
correspond to them ``retweeting'' a tweet from someone they follow. The above
probabilities were chosen to plausibly model the relative frequencies of
creating versus importing content; an experimenter may equally estimate their
values from a real dataset.

If a node $i$ chooses to create a new packet, a new ID is generated and that packet is
added to $i$'s list of content. Neighbors of $i$ may later choose to import this new
packet into their own content lists, letting it spread throughout the network.
Importing works as follows. If node $i$ chooses to import content, one of
$i$'s neighboring nodes $j$ is selected at random (assuming it has neighbors).
Once $j$ is selected, the information packets in $j$'s list are scanned from
most recently generated (or imported) to earliest generated (or imported). The
scan stops when an information packet is found that is not contained in $i$'s
list. If no such packet can be found, no action for node $i$ is taken and the
next activated node is considered. If such a packet is found, it is copied from
node $j$ to node $i$.

This process is repeated for the next node that has been activated during the
current time step. The simulation of transactions halts when $3000$ time steps
elapse. With the chosen values of $p_\mathrm{create}$ and $p_\mathrm{import}$, each
node will on average participate in the transaction model $1000$ times.
To avoid any pathological effects the nodes are activated in randomized
order for each time step.

Once the transactions have been simulated 
and we have a timeline of packets for each node, we then compute
the average time it takes for packets to flow between nodes. For every pair of
nodes in a graph, we computed the intersection between their respective sets
of information packets. We thus obtained each packet that both nodes in a pair
either imported or created. We then computed the average time $T_{ij}$ required
for packets to travel between nodes $i$ and $j$
\begin{equation}
T_{ij} = \sum_{k=1}^{n_{ij}} \left| t_{i}^{(k)} - t_{j}^{(k)} \right|,
\label{eqn:Tij}
\end{equation}
where $n_{ij}$ packets are shared by nodes $i$ and $j$, and $t_{i}^{(k)}$
indicates the time step at which packet $k$ was created (or arrived) at node
$i$. In order to remove noise resulting from small sample sizes, all $T_{ij}$
for which $n_{ij}<100$ were discarded. 
Note that we are not measuring a causal or directional relationship between
the node pair; a shared packet could easily have been created by a third node
and then eventually reached both $i$ and $j$ through the importing process.
The delay time $T_{ij}$ is a dynamical measure of closeness between the nodes.

\subsection{Symbolic Regression}
\label{subsec:symbolicregression}

Given a network topology and the information flow times $T_{ij}$
(Eq.~\eqref{eqn:Tij}), we then constructed a matrix $D$ to serve as the
dataset for training  models to predict $T_{ij}$ as a
function of the structural properties of nodes $i$ and $j$. Each pair of nodes
is allocated its own row. One column in $D$ contains the $T_{ij}$ values, while
the remaining columns correspond to structural network properties of node $i$,
node $j$, or some metric relating them.
An experimenter is free to choose which metrics to use. The individual node
properties we used were node degrees $k_i$, $k_j$; clustering coefficients
$c_i$, $c_j$; eccentricities $e_i$, $e_j$; node betweennesses $B_i$, $B_j$;
eigenvector centralities $x_i$, $x_j$, where $x_i$ is the $i$-th element of
the leading eigenvector of the network's adjacency matrix; and closeness
centralities $C_i$, $C_j$. For node-pair properties we used the length $L_{ij}$
of the shortest topological path between $i$ and $j$. Finally, we included
global network properties $N$, the number of nodes; $M$, the number of edges;
$r$, the degree-mixing assortativity coefficient~\cite{newman_mixing_2003}; and
the graph's diameter $\Delta$ and radius $\rho$. These global quantities were the same for all
rows of $D$, but providing them gives the optimization method a set of plausible constants to choose
from.\footnote{These can also become variables if one chooses to apply SR to a
dataset containing multiple networks of different sizes.}

We then perform symbolic regression (SR) on this dataset to find functions $f$
that predict $T_{ij}$ as a function of the node pair's structural
properties:
\begin{equation}
  \begin{aligned}
    T_{ij} = f \bigl( & k_i, c_i, e_i, B_i,x_i, C_i, \; k_j, c_j, e_j, B_j,x_j, C_j, \\
	                  & L_{ij}, N, M, r, \Delta, \rho \bigr).
  \end{aligned}
  \label{eqn:SRfunction}
\end{equation}
Symbolic regression performs model selection and parameter estimation
simultaneously to determine the functional form of Eq.~\eqref{eqn:SRfunction}. A
commonly-employed method for instantiating symbolic regression is genetic
programming~\cite{koza_genetic_1992}, a stochastic optimization method that simultaneously optimizes a
population of equations to increasingly fit the supplied data matrix $D$. As the name
implies, this method is loosely based on Darwinian evolution. An initial
population of random equations are assessed against $D$: models with
high error are discarded, while models with lower error are retained. The
now-vacant slots in the population are filled by repeatedly copying and mutating
a single equation, or producing two new equations by performing sexual
recombination with a pair of surviving equations. Mutations involve adding,
removing, or altering a term in the equation. 

The SR implementation we used in this study
incorporates multiobjective optimization to perform
search~\cite{bongard_automated_2007,schmidt_distilling_2009}. The errors and
sizes of the models in the population are computed. Size is defined as the total
number of operators and operands in the equation. The Pareto front of models
with least error and smallest size is determined, and models off this front are
discarded. New models are generated by randomly choosing surviving models on
the front. When run against a dataset generated by a single scale-free network
composed of $N=250$ nodes, the best equation found\footnote{Note that SR was prevented from using numerical prefactors to
enforce greater structural diversity in models along the Pareto front.}, in terms of balancing
complexity and accuracy, was
\begin{equation}
	T_{ij} = L_{ij}\left[1 + \ln\left(L_{ij} + k_i + k_j - c_j +  \frac{N - k_i^2 - k_j^2}{L_{ij}^N + k_i k_j - \rho} \right)\right].
\label{eqn:BA_SR}
\end{equation}
This equation achieved a high correlation coefficient of $R = 0.88$ when
compared with the simulated $T_{ij}$.
We remark that Eq.~\eqref{eqn:BA_SR} seems plausible in nature: the dominant variable
is the distance $L_{ij}$ between $i$ and $j$, which is intuitive for the
transaction model. The degrees of $i$ and $j$, the clustering of $j$ and global
network properties $N$ and the network radius then comprise a small, logarithmic
correction to $L_{ij}$.
Other variables did not factor into this function.

\subsection{Tampered networks}
\label{subsec:tamperednetworks}

To test the ability of the SR model to indicate the presence
of a hidden node, we need access to a ground truth test bed. To create such a
test using our model networks (Sec.~\ref{subsec:networkmodels}), we generate a
new scale-free network, simulate transactions on it
(Sec.~\ref{subsec:transactionmodel}), then choose one or more nodes to
\textbf{hide}; they are removed before computing the network structural metrics
and the information flow times (Eq.~\eqref{eqn:Tij}). In this way hidden nodes
fully participate in the flow of packets, but otherwise they are unknown to the
symbolically regressed model. Comparing the SR model's delay predictions
(Eq.~\eqref{eqn:BA_SR}) to the simulated delay times, we can \emph{a posteriori}
search for systematic errors among the neighbors of the hidden node or nodes.

To measure the effects of the hidden node we study three quantities. The first
is the coefficient of determination ${R}^2$ between the $T_{ij}$'s measured for the
non-hidden node-pairs from the transaction simulations and the predicted
$T_{ij}$'s from the SR model, where $R$ is the Pearson correlation coefficient.
If the value of $R^2$ drops significantly compared to $R^2$ for the untampered
network, then that supports the ability for us to detect missing or hidden
nodes.

Beyond this global measure we also use two local measurements to assess the
effect a hidden node has on a single non-hidden node $i$:
\begin{align}
	\mathrm{RMSE}_i = & \sqrt{\mathbb{E}_j\left[{\left(T_{ij}^\mathrm{pred} - T_{ij}^\mathrm{obs}\right)}^2\right]}, \label{eqn:rmse}  \\
	\mathrm{Bias}_i = & \mathbb{E}_j\left[T_{ij}^\mathrm{pred} - T_{ij}^\mathrm{obs}\right], \label{eqn:bias}
\end{align}
where the expectation $\mathbb{E}_j\left[\cdot\right]$ runs over all (non-hidden) nodes
$j \neq i$ that are connected to $i$ ($L_{ij} < \infty$), and
$T_{ij}^\mathrm{pred}$ and $T_{ij}^\mathrm{obs}$ denote the flow time predicted
by the SR model and the actual flow time observed from the simulations,
respectively.
The $\mathrm{RMSE}_i$ captures the magnitude of the SR model's error for node $i$,
while $\mathrm{Bias}_i$ measures whether it consistently
over- or under-estimated $T_{ij}$. A positive bias indicates that information is traveling
faster than expected by the SR model.

\section{Results}
\label{sec:results}

\begin{figure}[t!]
\centerline{\includegraphics[]{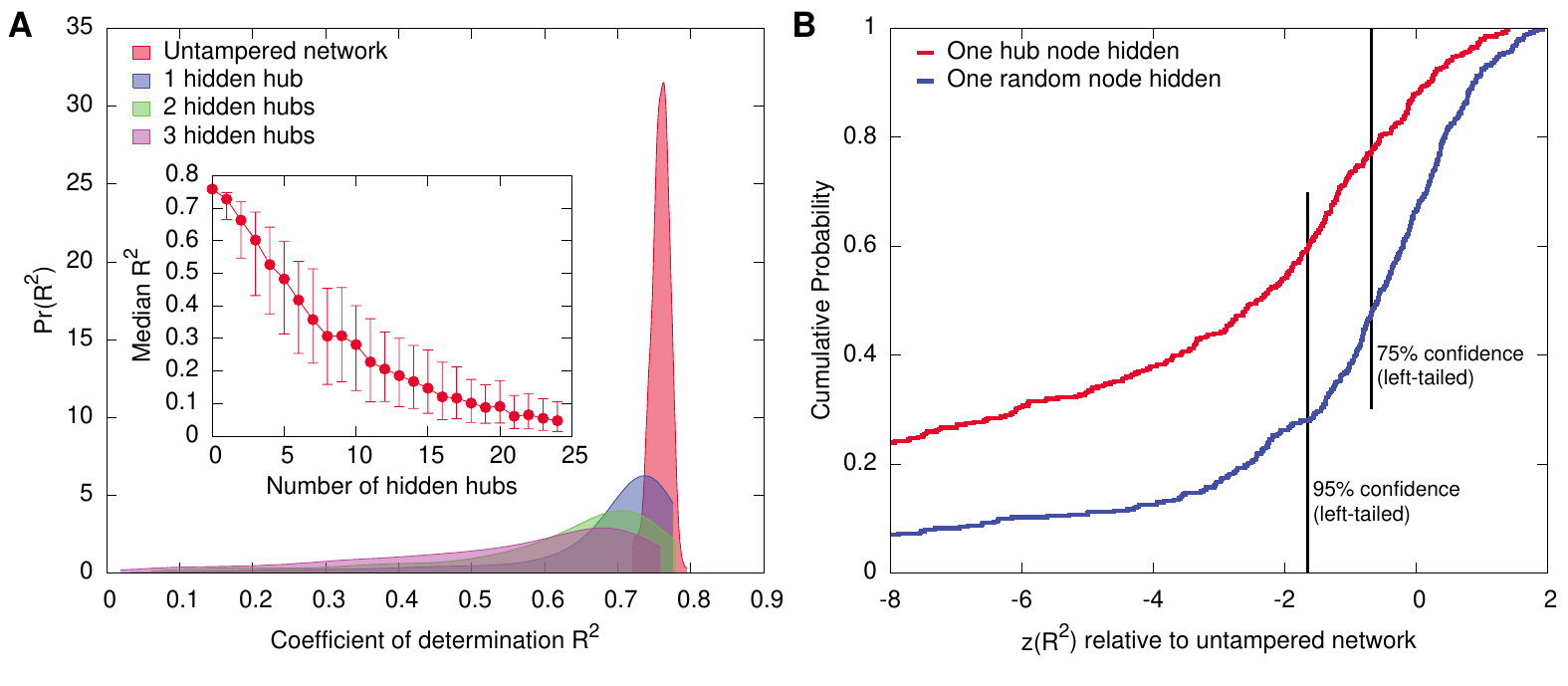}}
\caption{\label{fig:BA_vals}
    Detecting hidden nodes in scale-free networks.
    \lett{A} 
    The distribution of correlation coefficient comparing the
    simulated transaction times and those predicted by the symbolically
    regressed model. We see for networks drawn from the same ensemble as
    the training network (untampered) that $R^2$ is sharply peaked around 0.78.
    The distribution of $R^2$ changes significantly after a single node removal
	(Mann-Whitney U test $p < 10^{-10}$, Cohen's $d = -0.898$).
    (inset) The median $R^2$ decreases as more high-degree nodes are removed.
    \lett{B}
    The likelihood of detecting a single hidden node by the change in $R^2$.
    Using the distribution of $R^2$ for the untampered network as the null
    model, we standardize the $R^2$ distribution for networks with one hidden
    node. Looking at this distribution we see that nearly 60\% of the time we
    can successfully detect that a relatively high degree node is absent with
    95\% confidence. If we consider lower degree hidden nodes, which are more
    challenging to discover as they tend to participate less in information flow,
    this drops to approximately 25\%.
    Distributions shown in panel A were computed using kernel density
    estimation; each curve was truncated at the largest value observed
    in the ensemble data to indicate the empirical ranges of $R^2$.
}
\end{figure}

Our first experiment consisted of measuring the change in the
coefficient of determination $R^2$ for tampered scale-free networks
(Sec.~\ref{subsec:tamperednetworks}). To do this we first generated an ensemble of
100 untampered scale-free networks (Sec.~\ref{subsec:networkmodels}) and simulated
transactions on each (Sec.~\ref{subsec:transactionmodel}). 
We applied the SR model of $T_{ij}$ to these networks (Eq.~\eqref{eqn:BA_SR})
and computed $R^2$ for each. As shown in Fig.~\ref{fig:BA_vals}A, the distribution of
$R^2$ was sharply peaked around $R^2\approx 0.77$, the value that the SR
model achieved on its training data (Sec.~\ref{subsec:symbolicregression}).
The narrowness of this distribution indicates that the SR model has useful
predictive power.

Next we generated another ensemble of scale-free networks and simulated
transactions, but now we tampered with each network by hiding one random
hub\footnote{We take a hub to be a randomly chosen node that was introduced in
the first 20\% of the network growth process, taking advantage of preferential
attachment's early-mover-advantage.}. We see a significant drop in accuracy (lower $R^2$) for
the SR model on these tampered networks (Mann-Whitney U test $p < 10^{-10}$,
Cohen's $d = -0.898$), indicating that we are likely to see the effect of a
hidden node by a drop in the accuracy of the model. Hiding multiple hubs leads
to even greater losses in accuracy (Fig.~\ref{fig:BA_vals}A and inset).

However, the comparisons shown in Fig.~\ref{fig:BA_vals}A are for an ensemble
of networks, while practically we seek to detect the presence of a missing
node in a \emph{single} network. To determine if this is feasible we standardized
the distribution of $R^2$ for the ensemble of networks with a single hidden
hub relative to the untampered ensemble, giving a z-score $z(R^2)$ for each
tampered network. Large negative values of $z$ indicate a statistically significant
drop in $R^2$. The cumulative probability distribution shown in Fig.~\ref{fig:BA_vals}B
tells us that nearly 60\% of the tampered ensemble has $z(R^2) < -1.6449$,
meaning that nearly 60\% of the time we can determine with 95\% confidence that a
single network is missing a hub.
The 50\% confidence limit, $z < 0$, corresponding to how well we can beat a
coin-flip, is nearly 90\%.

\begin{figure}[t!]
\centerline{\includegraphics[]{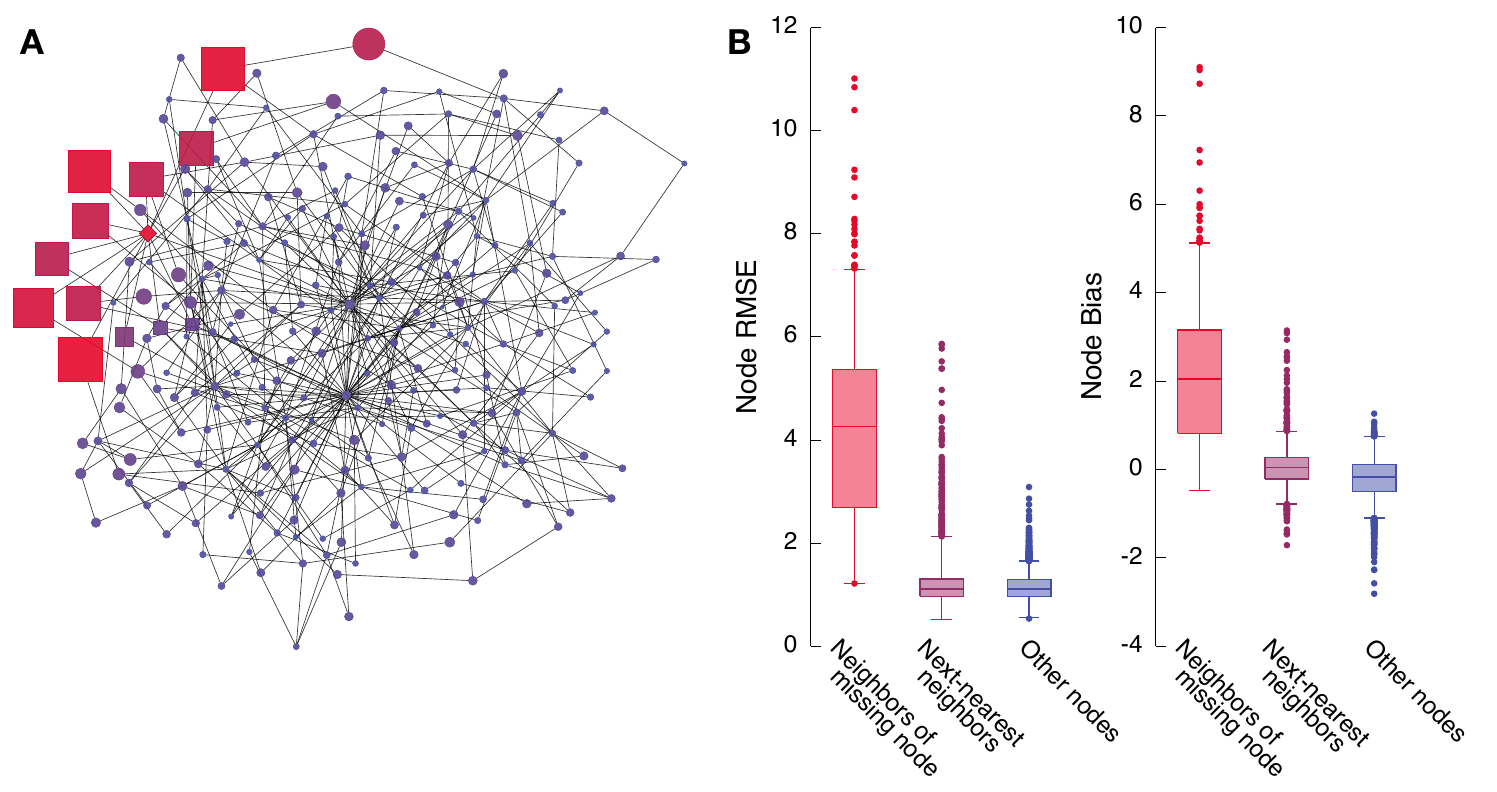}}
\caption{\label{fig:locationMissingNode}
    Identifying the location of a missing node.
    \lett{A}
    A scale-free network of 250 nodes with a single node hidden
    ($\diamond{}$). The neighbors of the hidden node are indicated with
    $\square{}$ while other nodes are $\circ{}$. The size and color of each
    node is proportional to the rms error of the information transfer
    time from that node to every other node in the network. We see that the
    neighbors of the missing node consistently have higher errors than the rest
    of the network.
    \lett{B}
    The distributions of error and bias across the ensemble of tampered
    networks for the hidden node's neighbors, next-nearest neighbors, and
    non-neighbors. The median error for neighbors is approximately $4.33$
    timesteps while for non-neighbors it is approximately $1.11$ timesteps.
    The distributions are significantly different (Mann-Whitney U test $p
    \ll 10^{-10}$, Cohen's $d=4.69$). The next-nearest neighbors have errors
    comparable to non-neighbors ($p = 0.052$) but we see a greater number of
    outliers skewing upward. This indicates that there are some network effects
    in how errors propagate, but they are relatively rare.
	Likewise, we see positive bias for neighbor nodes, significantly higher than for non-neighbors
	(Mann-Whitney U test $p \ll 10^{-10}$, Cohen's $d=3.37$). This positive bias
	indicates that information spreads faster from (or to) neighbors of the hidden node than
	the SR model expects, supporting the intuition behind Fig.~\ref{fig:cartoonTransactionsModel}
    To control for the centrality of the hidden node, in each realization the
    hidden node was the node with the fifth highest degree.
}
\end{figure}

These results indicate that the presence of a single hidden node can often be
detected. An important question, however, is whether or not we can identify
the \emph{location} of this hidden node. To study this, we computed the errors
and biases (Eqs.~\eqref{eqn:rmse} and~\eqref{eqn:bias}) of each node in a
tampered network. If the neighbors of the hidden node show significant error
or bias, then that means we can determine the location of the hidden node. We
show a network diagram of one tampered scale-free network in Fig.~\ref{fig:locationMissingNode}A. Node
color and size is proportional to $\mathrm{RMSE}$ and the hidden node is
indicated with a diamond ($\diamond{}$). We observe that many neighbors of the
hidden node have far greater $\mathrm{RMSE}$ than other nodes in the network.
This is exactly the evidence needed to estimate the hidden node's location
within the network topology.

To determine if these results are significant, we computed, for the ensemble
of scale-free networks with one hidden hub, distributions of $\mathrm{RMSE}$
and $\mathrm{Bias}$ separately for neighbors of the missing hub, next-nearest
neighbors, and other nodes.
The $\mathrm{RMSE}$ was significantly larger (Mann-Whitney U test $p \ll
10^{-10}$, Cohen's $d=4.69$) for neighbors of the hidden node across the
entire ensemble (the median error for neighbors was $\approx 4.33$ timesteps
compared with $1.1$ timesteps for other nodes).
Next-nearest neighbors, those nodes two steps away from the hidden node in the
original topology, did not show a significant change in error relative to other
nodes in the network ($p=0.052$). However, a number of outliers do overlap
with the $\mathrm{RMSE}$ values for the nearest neighbors, indicating that
longer-range network effects are rare but do occur.

At the same time, the $\mathrm{Bias}$ was also positively skewed for neighbors
of the hidden node (median $\mathrm{Bias} \approx 2.1$ timesteps), indicating that our
intuition from Fig.~\ref{fig:cartoonTransactionsModel} was correct. Next-nearest
neighbors have no discernible bias (median $\mathrm{Bias}\approx 0.03$), while other nodes
actually have a slightly negative bias (median $\mathrm{Bias}\approx -0.18$),
indicating the information in the rest of the network actually travels slightly
slower than expected due to the hidden node (however, a zero bias cannot be
ruled out for this group).

\section{Discussion}
 
We have shown that the presence of hidden nodes can be inferred by modeling how
network topology influences a dynamical process overlaying that network. We
focused on an idealized information flow dynamics but there is great potential
for applying this to other model dynamics. For future work, we intend to use
our methodology alongside real world data on information cascades and other
dynamical processes and to further study how different classes of network
topologies help or hinder the node discovery process. We also plan to better incorporate 
the \emph{directionality} of information flow, which was neglected here by the absolute value 
used in the equation for $T_{ij}$.

It is not particularly surprising that perturbing a network, which then leads to
perturbed metrics such as those used in Eq.~\ref{eqn:SRfunction}, will lead to
a reduction in the accuracy of an SR model (e.g., Eq.~\ref{eqn:BA_SR}). This was
shown in Fig.~\ref{fig:BA_vals}.
However, we have shown (Fig.~\ref{fig:locationMissingNode}) that the loss in
accuracy is \textbf{localized} and correlates with the position of the defect,
indicating that we are extracting useful information and not merely randomizing
the terms within the SR model's functional form.

More generally, looking for discrepancies in the speed of information flow (or
other quantities) can be used to study not just missing nodes but other defects
and errors, such as missing links or false links that incorrectly appear in the
network, false nodes that do not actually exist, the splitting of a true node
into multiple false nodes, or the merging of multiple true nodes into a single
false node. Some of these errors will likely prove more challenging to detect
than others, but the benchmarking procedure we have introduced here may offer some
hope towards tackling these problems.

\section*{Acknowledgments}

The authors are grateful for the computational resources provided by
the Vermont Advanced Computing Core which is supported by NASA (NNX 08A096G).
JCB was supported in part by NSF grant \#PECASE--0953837 and DARPA grant \#W911NF--11--1--0076. 
JCB and CMD were supported in part by DARPA grant \#FA8650--11--1--7155.
PSD was supported by NSF CAREER Award \#0846668.

\singlespacing{}

\end{document}